\documentclass[12pt,preprint]{aastex}
\usepackage{psfig}
\newcommand{\lta}{$\; \buildrel < \over \sim \;$}
\newcommand{\simlt}{\lower.5ex\hbox{\lta}}
\newcommand{\gta}{$\; \buildrel > \over \sim \;$}
\newcommand{\simgt}{\lower.5ex\hbox{\gta}}
\newcommand{\cm}{{\rm\,cm}}
\newcommand{\kms}{{\rm\,km\,s^{-1}}}
\newcommand{\msun}{{\rm\,M_\odot}}

\newcommand{\gs}{{\rm\,g\,s^{-1}}}

\newcommand{\pb}{\phi_{\scriptscriptstyle 98}}
\newcommand{\ps}{\phi_{\scriptscriptstyle 67}} 
\newcommand{\mo}{{\rm\,G\, cm^3}}

\slugcomment{Accepted for publication in ApJ}
\shortauthors{Belle, Howell, Sirk, \& Huber}
\shorttitle{EUVE observations of EX Hya}

\begin{document}

\title{An EUV Study of the Intermediate Polar EX Hydrae}

\author{Kunegunda E. Belle\altaffilmark{1,2},  
Steve B. Howell\altaffilmark{1,2}, Martin Sirk\altaffilmark{3}, and
Mark E. Huber\altaffilmark{1,2}}

\altaffiltext{1}{Astrophysics Group, Planetary Science
Institute, 620 N. 6th Ave.  Tucson, AZ 85705; keb@psi.edu,
howell@psi.edu, mhuber@psi.edu}
\altaffiltext{2}{Dept. of Physics and Astronomy, University of Wyoming,
Laramie, WY 82071}
\altaffiltext{3}{Space Sciences Laboratory, University of California, 
Berkeley, CA 94720; sirk@alma.ssl.berkeley.edu}

\begin{abstract}
On 2000 May 5, we began a large multi-wavelength campaign to study the
intermediate polar, EX Hydrae.  The simultaneous observations from six
satellites and four telescopes were centered around a one million
second observation with {\it EUVE}.  Although EX Hydrae has been
studied previously with {\it EUVE}, our higher signal-to-noise
observations present new results and challenge the current IP models.
Previously unseen dips in the light curve are reminiscent of the
stream dips seen in polar light curves.  Also of interest is the
temporal extent of the bulge dip; approximately 0.5 in phase, implying
that the bulge extends over half of the accretion disk.  We propose
that the magnetic field in EX Hydrae is strong enough (a few MG) to
begin pulling material directly from the outer edge of the disk,
thereby forming a large accretion curtain which would produce a very
broad bulge dip.  This would also result in magnetically controlled
accretion streams originating from the outer edge of the disk.  We
also present a period analysis of the photometric data which shows
numerous beat frequencies with strong power and also intermittent and
wandering frequencies, an indication that physical conditions within
EX Hya changed over the course of the observation.  Iron spectral line
ratios give a temperature of $\log\,T=6.5-6.9$ K for all spin phases
and a poorly constrained density of $n_{\rm
e}=10^{10}-10^{11}\cm^{-3}$ for the emitting plasma.  This paper is
the first in a series detailing our results from this multi-wavelength
observational campaign.
\end{abstract}


\section{INTRODUCTION}\label{intro}
Intermediate polars (IPs) have been the focus of numerous studies
since they first emerged as a separate class of cataclysmic variables
(CVs).  Possessing observational characteristics of both dwarf novae
and magnetic cataclysmic variables, the current model for intermediate
polars paints them as a fusion of these two types of systems.  An
asynchronously rotating magnetic white dwarf primary accretes
magneto-hydrodynamically controlled material from a truncated
accretion disk which is fed via an accretion stream by a red dwarf
secondary star \citep{pat94}.  The magnetically controlled accretion
produces observational properties similar to those of a polar, while
the visible accretion disk creates its signature double peaked
spectral lines and other typical CV properties.  The different
components of the intermediate polar system: the secondary star,
accretion stream, accretion disk, accretion region on the white dwarf
surface, and the white dwarf itself, separately dominate in wavelength
regions ranging from infrared to X-ray.  While this makes IPs very
complicated objects to study, it also provides the opportunity to
examine a multitude of astrophysical phenomena: from hot plasmas and
accretion processes in the presence of high gravity and temperature to
stellar astrophysics and binary evolution.

An opportunity for an in-depth observational study of an intermediate
polar arose when we were awarded one million seconds of time on the
{\it Extreme Ultraviolet Explorer Satellite} ({\it EUVE}) to observe
EX Hydrae.  In collaboration with other researchers, we used a number
of other satellites to obtain simultaneous observations during the
{\it EUVE} observation.  X-ray data were obtained with {\it Chandra}
(spectroscopy), the {\it Rossi X-ray Timing Explorer} ({\it RXTE},
spectroscopy and photometry), and the Unconventional Stellar Aspect
(USA) Experiment (photometry).  Along with {\it EUVE}, ultraviolet
data were taken with the {\it Far Ultraviolet Spectroscopic Explore}
({\it FUSE}, spectroscopy) and the {\it Hubble Space Telescope} ({\it
HST}, spectroscopy).  Several ground based observatories were used to
obtain optical data: Apache Point Observatory (APO, spectroscopy),
Rosemary Hill Observatory (RHO, photometry), and Braeside Observatory
(BO, photometry).  Infrared spectra were obtained at the United
Kingdom Infrared Telescope (UKIRT).  This paper is the first in a
series which will detail the observations obtained during our
campaign.

\section{REVIEW OF EX HYA IN THE EUV}
EX Hya has been the object of many previous observational studies for
two main reasons: it is very bright ($V\sim13$) and, with an
inclination of $i=77^{\circ}\pm1^{\circ}$
\citep{hellier87,siegel89,allan98}, EX Hya is one of only two X-ray
eclipsing intermediate polars (the other is XY Ari, Hellier 1997).
However, some difficulties arise when interpreting observational data
due to the fact that the relatively long white dwarf spin period of
$67.027$ minutes is roughly two-thirds that of the relatively short
binary orbital period of $98.257$ minutes \citep{kait87}.  Masses for
the white dwarf in EX Hya have been determined using various methods
and have yielded a range of values: measurements of the accretion
stream shock height and temperature give $M_{\rm WD}\sim0.5M_{\odot}$
\citep{fuj97,allan98}, radial velocities derived from optical time
resolved spectra have led to slightly higher values for the white
dwarf mass, $0.78M_{\odot}$ \citep{hellier87}, and a mass as high as
$0.85M_{\odot}$ has been calculated using binary star parameters
\citep{hur97}.

The accretion curtain theory of IPs, developed from high energy
studies, suggests that modulation observed on the white dwarf spin
period is due to varying optical depths through curtains of accreting
material along our line of sight to the emission regions.  However,
according to recent X-ray observations
\citep{beuer88,singh93,allan98}, it is more likely that photoelectric
absorption combines with an occultation of one of the accretion poles
to give the observed spin modulation.  Maximum (minimum) of the spin
phase in EX Hya occurs when the upper magnetic pole points away from
(toward) the observer.

Prior to this campaign, the most detailed EUV study of EX Hya was that
performed by \citet{hur97}, who obtained 180 ks of data with the {\it
EUVE} satellite.  They found the spin phased light curve to be
sinusoidal in shape and modulated by a factor of $\sim3.7$, with a
maximum count rate of $\sim0.30\,{\rm s}^{-1}$ occurring at $\phi
\sim0.05$.  The binary phased light curve exhibits a bulge eclipse
which extends over $\phi\sim0.6-1.0$ \citep{mauche99}.  This bulge
eclipse has a non-zero flux, being $\sim0.17$ of the non-bulge
eclipsed flux averaged over the white dwarf spin phases, and has an
intrinsic column density determined to be $1.3\times10^{20}\cm^{-2}$
\citep{hur97}.  There is also a feature attributed to an eclipse by
the secondary star at $\phi=0.97$ \citep{mauche99} with a centroid
which remains constant to within $6$ s for all white dwarf phases
implying an emission region located within $\pm0.5 R_{\rm WD}$ of the
white dwarf.  The ingress and egress times of the secondary eclipse
were used to constrain the EUV emitting region size to be $<0.23 R_{\rm
WD}$ at 68\% confidence \citep{hur97}.

The EUV spectra presented in \citet{hur97} contain highly ionized iron
and neon emission lines which are indicative of a $10^7$ K plasma and
a volume emission measure of $3\times10^{54}\cm^{-3}$.  Using the
volume emission measure and the characteristic cooling time for a
$10^7$ K plasma, they determined the accretion rate onto the white
dwarf to be $\dot{M}_{\rm WD}\sim 3\times10^{16}\gs$.

\section{{\it EUVE} OBSERVATIONS}
EX Hya was observed with the {\it EUVE} deep survey (DS; $67-364$\AA)
instrument and short wavelength spectrometer (SW; $70-190$\AA,
$\Delta\lambda\sim0.5$\AA, Bowyer 1995) continuously for six weeks
from 2000 May 5 - June 10, totaling approximately one million seconds
of on target time given a $\sim30\%$ duty cycle.  In spite of two safe
holds caused by a battery failure and very high He II $\lambda304$\AA\
airglow, the data quality is very good.  About $10\%$ of the DS
photometer data and a bit less of the SW spectrometer data were
rejected during periods of high background/airglow since the deadtime
corrections became very large ($200-700\%$).  Typical DS deadtime
corrections for the remaining data were around $115\%$ (i.e., the
detector was too busy to process new photons $15\%$ of the time).  Two
types of products were produced: DS light curves and SW spectrometer
spectra, and each is discussed below in detail.

\subsection{Photometry}\label{phot} 
\begin{figure}
\begin{center}
\epsscale{0.9}
\plotone{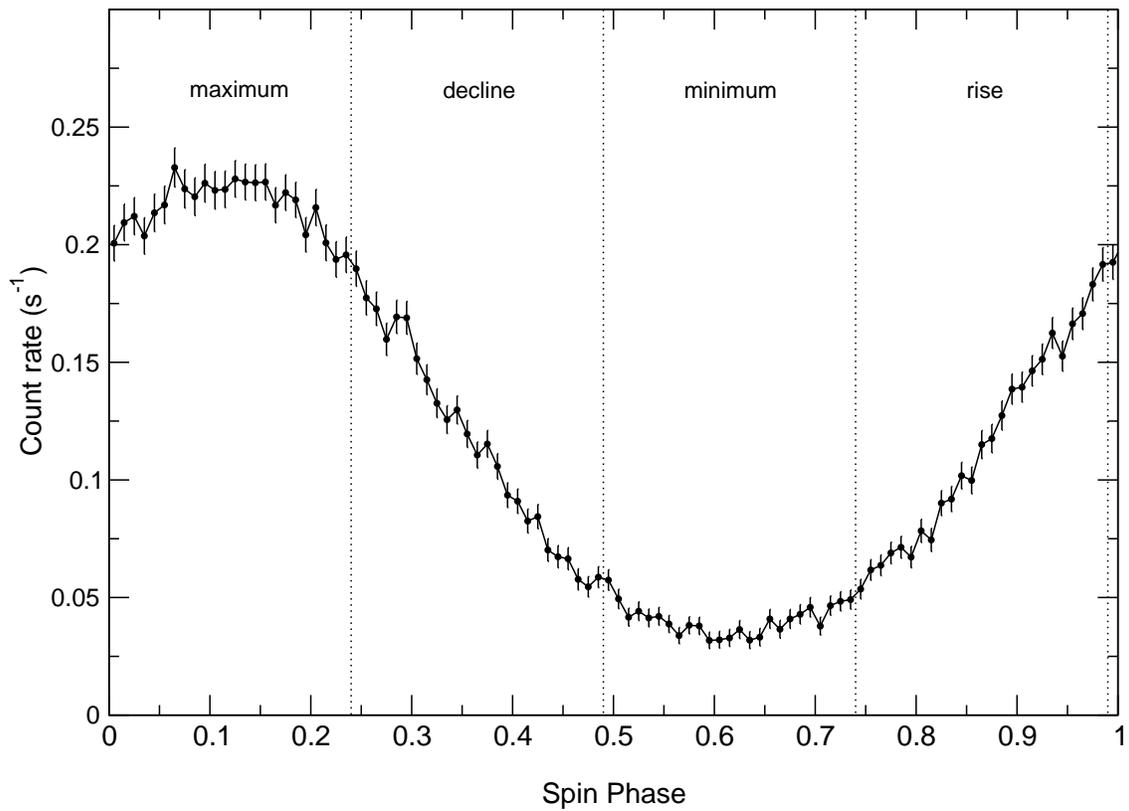}
\figcaption[]{Summed EUVE spin phased light curve.  This light
curve has a resolution of $\Delta\ps=0.01$ or $\sim40$ s. 
Spin phase delimitations are labeled.\label{splc}}
\end{center}
\end{figure}

\begin{figure}
\begin{center}
\epsscale{0.9}
\plotone{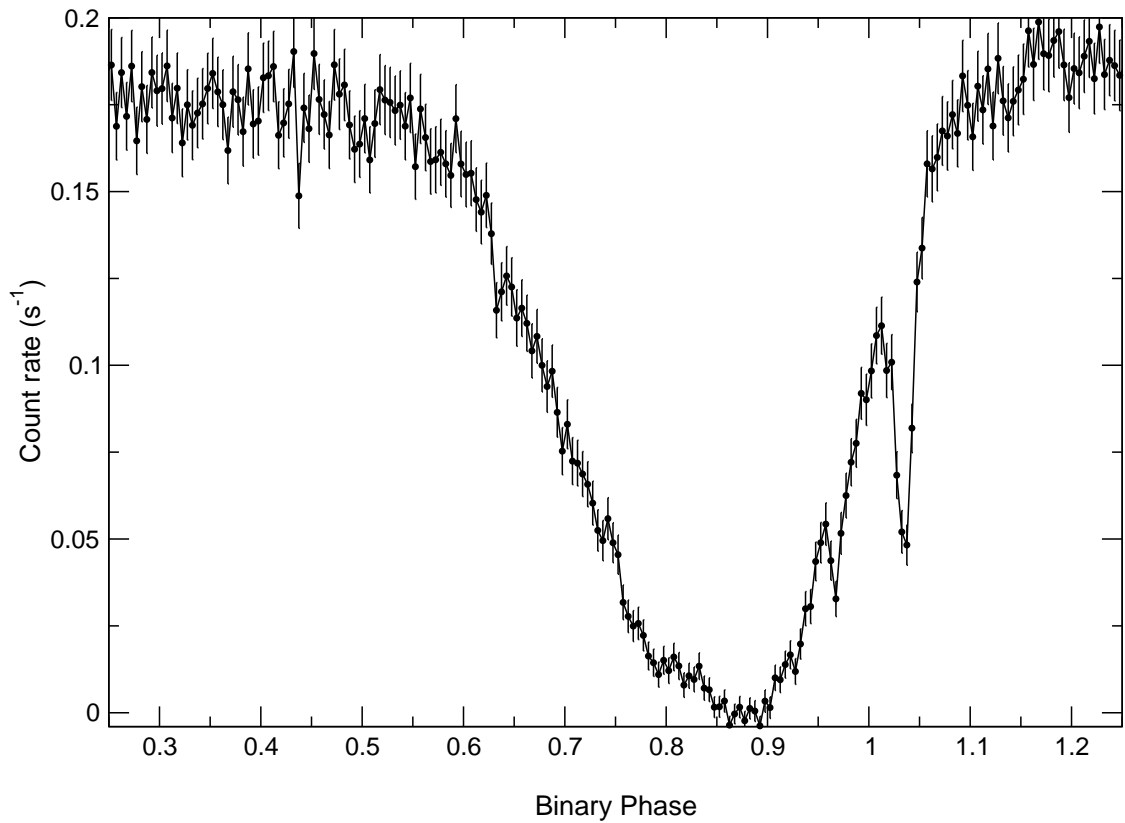}
\figcaption[]{Summed binary phased EUVE light curve.  The resolution
is $\Delta\pb=0.005$ or $\sim30$~s.  \label{bplc}}
\end{center}
\end{figure}

Photometry was performed from the DS images using a circular region of
radius 13 pixels centered on the source and an annular background
region with an inner radius of 50 pixels and outer radius of 100
pixels (1 pixel = $4\farcs6$, Sirk et al. 1997).  The photon counts
were then binned on a time interval of 10 seconds and barycentric
corrections were applied to the mid-time of each bin.  From this
photometric data set, phase folded light curves were created for the
binary period, the white dwarf spin period, and for a few combinations
of these two periods (see \S\ref{complcs}).  Figure \ref{splc} shows
the {\it EUVE} photometric data folded on the white dwarf spin
ephemeris, $T=2437699.8914(5) + 0.046546504(9)\mathrm E - 7.9(4)\times
10^{-13} \mathrm E^2$ \citep{ephem}, while Figure \ref{bplc} presents
the {\it EUVE} photometric data folded on the binary ephemeris,
$T=2437699.94179 + 0.068233846(4)\mathrm E$ \citep{ephem}.  The binary
ephemeris phase zero is based on optical binary eclipse timings, while
spin phase zero is based on optical spin maxima timings.  We will
denote the spin phase as $\ps$ and the binary phase as $\pb$.

The spin phased light curve (Figure \ref{splc}), with resolution
$\Delta\ps=0.01$ or $\sim40$ s, can be fit with the sinusoidal
function, $A + B \sin2\pi (\ps - \phi_0)$, where $A=0.122\pm0.001$
counts s$^{-1}$, $B=0.097\pm0.001$ counts s$^{-1}$, and
$\phi_0=0.865\pm0.001$.  Maximum occurs at $\ps=0.115\pm0.001$ and
minimum occurs at $\ps=0.615\pm0.001$; both values differ from the
expected values of $\ps=0.0$ (maximum) and $\ps=0.5$ (minimum), and
the maximum phase also differs from a previously determined value of
$\ps=0.040\pm 0.002$ \citep{mauche99}.  The shift in maximum phase is
real as the errors quoted for the spin ephemeris and period
\citep{ephem} combine to give an expected error of $\pm0.03\ps$ within
the light curve; indicating a possible change in accretion geometry.
The count rate at spin maximum is $\sim7\times$ the spin minimum count
rate; the non-zero minimum count rate value
($0.035\,\mathrm{counts\,s^{-1}}$) implies that either part of the EUV
emitting region is still in view during the minimum phases, or that
there is EUV flux within the binary system not confined to a small
occulted region.

Two theories have been advanced to explain the spin modulation in
intermediate polars: absorption by an accretion curtain
\citep{hellier87} and occultation of the EUV emitting region by the
white dwarf \citep{rosen91}.  While both theories often seem to
explain the observed phenomena in different IPs, the EUV flux in EX
Hya at minimum is less than half that at maximum, suggesting that if
one of the poles is occulted, the other may be partially absorbed or
the poles may have very different magnetic field strengths (which
could manifest as accretion poles with different temperatures.)  For
reference, we refer to phases $\ps=0.99-0.24$ as spin maximum, phases
$\ps=0.24-0.49$ correspond to spin decline, phases $\ps=0.49-0.74$ are
spin minimum, and phases $\ps=0.74-0.99$ are spin rise, as illustrated
in Figure \ref{splc}.

EUV modulation seen in the binary light curve (Figure \ref{bplc}, with
resolution $\Delta\pb=0.005$ or $\sim30$ s) is the result of the
inclination angle of EX Hya, rather than emission over the binary
orbit.  The fluctuations and dips seen throughout the light curve are
due to eclipses of varying natures; the large broad dip, or bulge dip,
centered at $\pb\sim0.85$, is caused by an increase in optical depth
along our line of sight to the EUV emitting region as viewed through
the bulge created by the impact of the accretion stream with the outer
edge of the accretion disk.  The temporal extent of this dip, from
$\pb=0.55-1.1$, implies that the bulge physically extends over half of
the accretion disk.  The dip at $\pb=0.97$ has generally been
attributed to an eclipse of the EUV emitting region by the white
dwarf, while the dip at $\pb=1.04$ has not been observed before in
EUV or other high energy photometry of EX Hya (e.g., Rosen et
al. 1991; Hurwitz et al. 1997; Mauche 1999).

\subsection{Spectroscopy}
\begin{figure}
\begin{center}
%\epsscale{0.8} 
\plotone{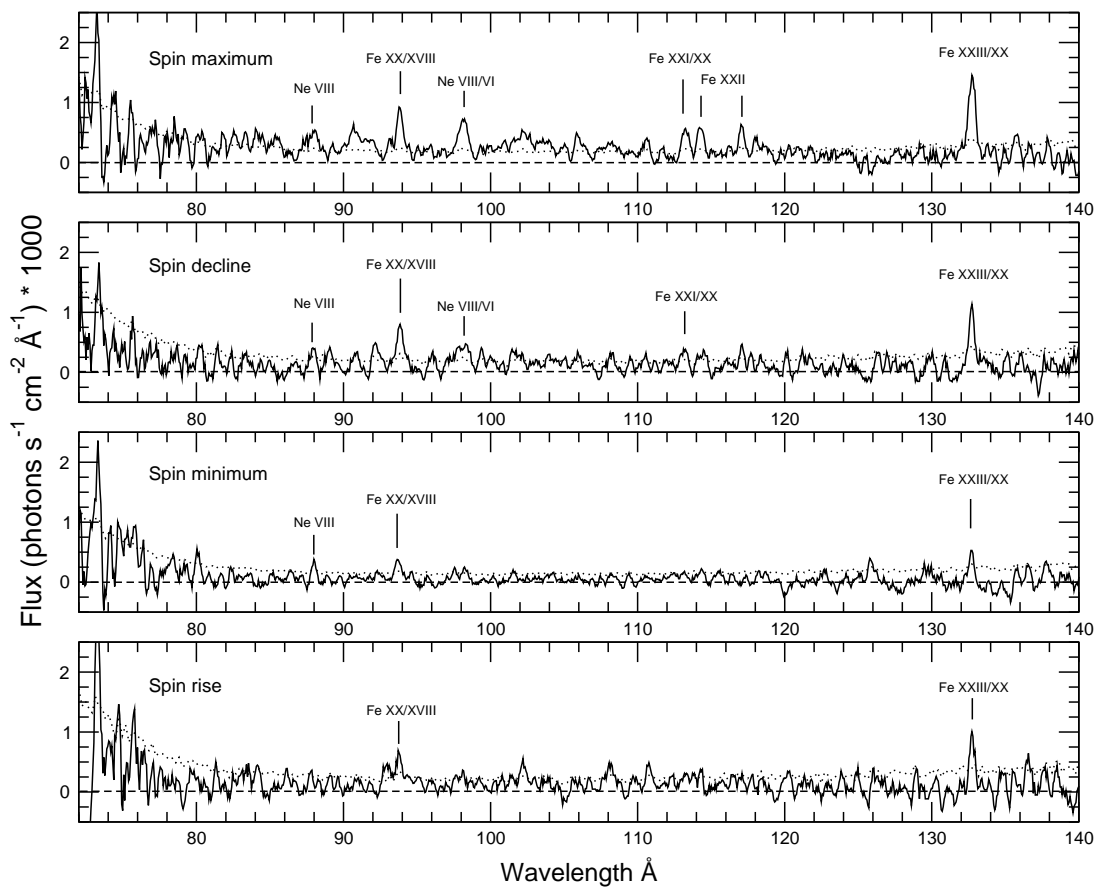} 
\figcaption[]{{\it EUVE} SW spectra combined on four spin
phases.  From top to bottom they are spin maximum, decline, minimum,
and rise.  Some of the stronger lines from Table \ref{t2} have been
labeled in this plot.  See Table \ref{t2} for the complete list of
identified lines.  The dotted line represents the $1\sigma$ error for
each spectrum.\label{spec}}
\end{center}
\end{figure}

To extract the {\it EUVE} SW spectra, a spectral region 13 pixels tall
centered on the spectrum, and two background strips each 100 pixels
tall were chosen (1 pixel = $4\farcs4$, Abbott et al. 1996).  The
background spectrum was fit with a smooth Legendre polynomial before
subtracting.  Utilizing this method ensures that no additional Poisson
noise is added to the spectrum \citep{hur97}.  Spectra were created
for a variety of binary and white dwarf spin phases.  Figure
\ref{spec} presents spin phased spectra, but see Figure 2 of Belle,
Howell, \& Sirk (2002b) for binary phased spectra.  The EUV spectra
presented here are not significantly better in signal-to-noise per
spectrum than those presented in \citet{hur97}, even though our
observations were $5.5\times$ longer.  Assuming equal background
levels during both observations, the signal-to-noise ratio in each of
our {\it four} phase bins should be roughly $2\times$ greater than the
signal-to-noise ratio in each of the two phase bins shown in
\citet{hur97}.  However, due to higher solar activity during our
observations which increased the {\it EUVE} background level, the gain
in signal-to-noise per spectrum is marginal.

The spin maximum spectrum exhibits the greatest flux, both in the
continuum and emission lines, while the spin minimum spectrum has a
continuum level consistent with zero.  The spin decline and spin rise
spectra have average flux levels that are roughly equal to each other,
in between the flux levels of the spin maximum and spin minimum
spectra.  Discerning between a continuum level and a forest of low
signal-to-noise emission lines is difficult for these spectra, thus we
only identify lines as probable that extend above the $2\sigma$ error
level (adopting a $3\sigma$ criterion for the lines virtually rules
out all line detections.)  Table \ref{t2} gives the observed
wavelengths, fluxes, and the log of the temperature at which the
emission line has its greatest power ($T_{\rm max}$) for the $2\sigma$
lines identified in the four spin phased spectra along with a few
other unblended lines which are used for density determinations (see
\S\ref{sect_spec}).  Gaussians were fit to the emission lines in order
to determine line parameters and the continuum was taken to be zero in
each of the four spectra.  The error reported for each $\lambda_{\rm
obs}$ measurement is a $1\sigma$ error and does not include the rms
error of the SW spectrometer, $0.11$\AA.  To aid us in the selection
of lines, each listed line was also identified in the summed spectrum.
The SW spectrometer has a moderate wavelength resolution of 0.5\AA,
causing many of the emission lines to be blended together.  For the
lines we believe are blended, both lines are listed in Table \ref{t2}
for the identification at a single wavelength measurement.  Emission
lines appearing in the EUV spectra of EX Hya are due to highly ionized
iron and neon and have formation temperatures of $10^{5.7}<T_{\rm
max}<10^{7.1}$ K.

\section{DISCUSSION}
\subsection{The Binary Phased Light Curve and Understanding the Binary 
Ephemeris}
The exact interpretation of the phenomena present in the EUV binary
light curve of EX Hya is dependent upon the accuracy of the binary
ephemeris.  Previous EUV and high energy binary light curves of EX Hya
have shown a dip at $\pb=0.97$ (or $\pb=0.98$, Mukai et al. 1998)
which has been attributed to a stellar eclipse of the EUV emitting
region or white dwarf \citep{rosen88,mauche99}.  However, there is no
direct evidence for a stellar nature of this dip, other than the dip
occurs {\em close} to the expected phase of $\pb=0.0$.  With a new dip
appearing at $\pb=1.04$, similar in nature to the $\pb=0.97$ dip, we
are lead to question which, if any, dip is stellar in nature, what
could be the cause of a non-stellar dip, and why the dip at $\pb=1.04$
has not been observed before.

The binary phasing, $T=2437699.94179+0.068233846(4)\mathrm E$, first
determined by \citet{mum67} (ephemeris constants were updated by
Hellier \& Sproats 1992), and still used today, is based on $U$ and
$B$ optical light photometric eclipse timings.  We note that the error
reported in the \citet{mum67} phasing propagates to only $\pm0.01\pb$
today.  Photometric eclipses seen in optical light in cataclysmic
variables with accretion disks are often due to eclipses of the hot
spot by the secondary star.  The optical eclipse depth in EX Hya was
seen to be shallow and variable over time \citep{hellier87} and during
outburst becomes broad and shallow \citep{reinsch90} implying that the
accretion disk is eclipsed rather than the white dwarf.  Thus,
photometric phase zero according to the \citet{mum67} binary ephemeris
for EX Hya is likely to be a hot spot eclipse and not true binary zero
(i.e., inferior conjunction of the secondary star).

How does the binary ephemeris photometric phase zero compare with
spectroscopic phase zero and true binary zero?  A stellar eclipse of
the hot spot (or bulge) denotes photometric phase zero, while
spectroscopic phase zero is taken to occur at the red-to-blue crossing
of the optical emission line radial velocity curve; the latter
theoretically occurring at inferior conjunction of the secondary star.
However, due to a bias in the (disk) emission lines from the hot spot,
the red-to-blue crossing can occur approximately when the hot spot is
moving perpendicular to the observer's line of sight.  For example,
\citet{mason00} showed that for cataclysmic variables with a strong
optical emitting hot spot, a phase offset exists between spectroscopic
and photometric phase zero; a stellar eclipse of the hot spot would
occur prior to the hot spot being perpendicular to our line of sight.
The measured radial velocity curves of the H$\beta$ and H$\gamma$
optical emission lines of EX Hya exhibit a red-to-blue crossing which
lags photometric binary phase zero by approximately $0.035$ in phase
\citep{hellier87}.  The optical emission lines are likely to be biased
by the motion of the hot spot \citep{hellier87} so that the
red-to-blue crossing may be interpreted as occurring when the hot spot
is traveling approximately perpendicular to the observer's line of
sight.  Thus, traveling along the binary orbit, true binary phase zero
would occur first, followed by photometric phase zero, and then
spectroscopic phase zero, i.e., an eclipse of the white dwarf by the
secondary star (if present) would be seen before photometric phase
zero.

Applied to EX Hya, this implies that an eclipse of the white dwarf by
the secondary would be seen at a phase prior to photometric binary
zero, such as $\pb=0.97$.  However, the minimum depth of a stellar
eclipse is likely to remain constant and not be dependent upon
the spin phase during which it is observed.  We will show in the
following section that a constant minimum is not the case for this
``eclipse''.  Therefore, although the phasing of the dip at $\pb=0.97$
may be correct for a stellar ``eclipse'', its classification as such
is still questionable.

\subsection{Comparative Light Curves}\label{complcs} 
\begin{figure}
\begin{center}
%\epsscale{0.8} 
\plotone{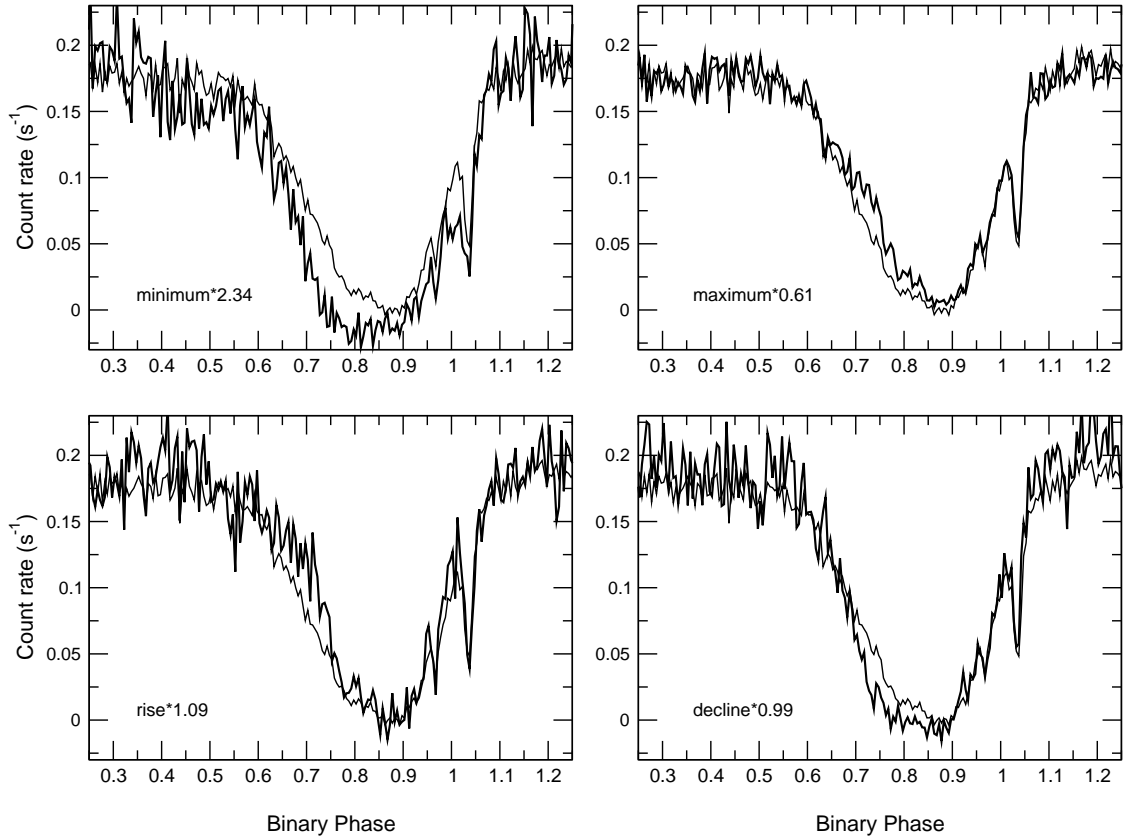} 
\figcaption[]{Plotted here are the data sets from the four spin
phases, phased over the binary orbital period.  Top left to right:
spin minimum and spin maximum; bottom left to right: spin rise and
spin decline.  In each graph, the thick line represents the individual
spin phase which has been scaled, for comparative purposes, to the
mean level of the summed light curve (thin line) by the number
indicated.
\label{wdcompare}}
\end{center}
\end{figure}

The identification of the origin of the dips seen at $\pb=0.97$ and
$\pb=1.04$ is facilitated by examining the binary phased light curve
extracted at different spin phases (Figure \ref{wdcompare}).  An
intriguing feature of the EUV light curves is the relative intensity
of the dips at phases $\pb=0.97$ and $\pb=1.04$.  Table
\ref{countrate} provides the unscaled mid-dip count rates for both the
$\pb=0.97$ and $\pb=1.04$ dips.  The $\pb=0.97$ dip is shallowest
during the spin maximum phase and not noticeably different from the
other small dips appearing throughout these light curves.  The count
rate at dip minimum in the spin maximum light curve ranges from
$\sim20\times$ greater than the mid-dip count rate in the spin minimum
light curve to $\sim2\times$ greater than the mid-dip count rate in
the spin decline light curve.  The difference between mid-dip count
rates for the $\pb=1.04$ dip is not as extreme, with the spin maximum
mid-dip count rate being $\sim8\times$ greater than the spin minimum
mid-dip count rate, ranging to $\sim2\times$ greater than the spin
decline value.

Gaussians were fit to the $\pb=0.97$ and $\pb=1.04$ dips in the
averaged and all four spin phased binary light curves.  The results of
the Gaussian fits are given in Table \ref{times}.  The errors on the
measured midpoint values fall within the light curve resolution of
$0.005\pb$.  The midpoint of the $\pb=0.97$ dip appears to be
consistent between the spin maximum and spin rise phases, at
$\pb=0.969$, and between the spin minimum and decline phases at
$\pb=0.972$, however, within the resolution of $0.005\pb$, these
values are the same.  The $\pb=1.04$ dip midpoint is also consistent
in phase at $\pb=1.038$.  FWHM and FWZI values are a bit more
scattered.  The FWHM values for both dips are the same, within errors,
between spin maximum and minimum.  The spin rise and decline FWHM
values are scattered and show no coherent pattern as are the FWZI
values for each dip over all spin phases.  The short duration of both
dips implies that the region being eclipsed is most likely small in
size and close to the white dwarf surface, i.e. the inner accretion
curtain close to the accretion pole where the EUV radiation is
emitted.

Other asymmetries between the binary EUV light curves of EX Hya in
Figure \ref{wdcompare} are seen within the broad dip extending from
$\pb\approx 0.55-1.1$, which is attributed to photoelectric absorption
of the accretion region by the bulge or hot spot \citep{cord85}.  This
broad dip feature has been seen in the light curves of previous EUV
and X-ray studies of EX Hya \citep{cord85,rosen88,hur97}.
\citet{rosen88} show a broad dip which extends for approximately
$\pb\sim0.7-0.97$ in the $0.02-2.5$ keV band and has an asymmetric
shape with a gradual decline and rapid rise while the eclipse seen in
\citet{cord85} ($0.04-2.0$ keV) is a little wider in phase.  The EUV
bulge dip as shown in \citet{hur97} and \citet{mauche99} extends from
$\pb\approx0.65-1.02$ and shows the same gradual decline and rapid
rise as the binary light curve presented in this paper.

In a typical cataclysmic variable or intermediate polar system with an
accretion disk, the hot spot, the region where the mass transfer
stream impacts with the accretion disk, is centered around binary
phase $0.8-0.9$ and does not extend ``backwards'' past the line of
centers.  Two things make the bulge in EX Hya atypical; it physically
extends over half of the outer edge of the accretion disk (see also
Hellier et al. 2000 who present a $B$-band light curve of EX Hya with
an orbital hump extending from $\pb\sim0.6-1.15$), and it extends past
binary phase zero.  However, we do note that some non-magnetic
cataclysmic variables also exhibit an extended hot spot region (e.g.,
Z Cha, Wood et al. 1986).

The bulge dip appears slightly different when compared between each
spin phase as shown in Figure \ref{wdcompare}.  Several differences
between the bulge dips in the different light curves are immediately
obvious: for example, during the maximum phase, the bulge dip is
shallower and has a more gradual decline.  However, the temporal
extent of the bulge dip appears to be the same in all spin phases,
with the exception of an extra shallow dip, preceding the bulge dip,
from $\pb\sim0.4-0.6$ in the spin minimum light curve.  One could
argue that the reason for the more rapid decline in flux in the spin
minimum light curve is simply due to the fact that there is already a
higher optical depth along the line of sight (the accretion curtain)
to the accretion pole during the minimum phase.  However, this
absorption effect would act to absorb the emission by the same amount
throughout the entire binary orbit rather than causing extra
absorption at a particular binary phase.  In the early phases of the
bulge dip, $\pb\sim0.65-0.7$, the spin maximum count rate is
$\sim12\times$ greater than the spin minimum count rate.  Comparing
this value to the count rate difference in the averaged spin light
curve ($\sim7\times$, Figure \ref{splc}), it is obvious that there
must be extra absorption in addition to the accretion curtains during
the bulge dip phases.

A change in bulge size is likely to manifest itself observationally in
wavebands additional to the EUV, and indeed, we do see a difference
when comparing our optical data to the optical light curves from
previous studies.  For example, our $B$-band light curve compared with
the $B$ light curve of \citet{mum67} shows a decreased eclipse depth
and very prevalent optical sidebands (see Figure 7 of Belle et
al. 2002b for a comparison of the light curves).  Speculating that
these sidebands might be seen in the EUV photometry of EX Hya, we
phased our EUV photometric data on common optical orbital sidebands
(or beat frequencies) usually seen in intermediate polars:
$\omega\pm\Omega$, $\omega\pm2\Omega$, and $2\Omega$, where $\omega$
is the spin frequency and $\Omega$ is the binary orbital frequency.
Coherent modulation, to some degree, was seen on all of these beat
frequencies.

In order to investigate the modulations seen on the orbital sidebands
and other periods, we searched our EUV photometry for periods using
the phase dispersion minimization (PDM) routine \citep{stell78}.  PDM
provides statistical testing for a given period via a two-sided
F-test.  Figure \ref{power} presents the results of this analysis, in
the form of a thetagram, with the y-axis representing $\Theta$ which
expresses frequency significance, with lower values of $\Theta$
indicating more statistically significant frequencies.  The
photometric data was searched for periodicity over time intervals of
10 min (144 cycles day$^{-1}$) to 20 days (0.05 cycles day$^{-1}$),
consistent with the available sampling window provided by our long
data set and the Nyquist sampling limit.  For the data set shown in
Figure \ref{power}, which contains 97,274 samples, a value of
$\Theta=0.976$ defines the 99\% confidence level for period
believability.  The 95\% confidence level is near $\Theta=0.99$.  The
occurrence of sampling aliases was lowered by increasing the bin size
over which the data were searched, as recommended by \citet{stell78}.

We see that in Figure \ref{power} essentially all the periods found
are real, they are very well defined, and we have identified a few
frequencies of interest on the figure, including that for the {\it
EUVE} satellite orbital period ($P_{\rm sat}=93.73$
min).\footnote{The nominal satellite period was 94.78 min, however,
during this observation, the period varied between 93.86 and 93.60
minutes.  We would like to thank the anonymous referee for calling
this to our attention.}  There are approximately 50 frequencies with
$\Theta$ values above the 95\% confidence level ($\Theta=0.99$)
appearing in the thetagram.  We have identified all of these
frequencies as beats between the spin, orbit, and satellite
frequencies, but have only labeled the more common beat frequencies on
Figure \ref{power}, and frequencies that have been identified in
previous photometry of EX Hya.  The $\Omega/2=7.33$ day$^{-1}$
frequency is strong with $\Theta=0.92$ and many of the frequencies
appearing in the range $0-10$ day$^{-1}$ are satellite beat
frequencies, but we have not labeled these frequencies in order to
avoid confusion on the plot.  The strongest beat frequencies are those
between the satellite and orbit or spin frequency, rather than between
the orbit and spin frequencies.  We note that the orbital frequency,
$\Omega$, is much stronger in the EUV than in our optical and X-ray
data ($\Omega$ and $\omega$ have roughly equal strengths in the EUV,
while $\omega$ is $\sim2\times$ stronger than $\Omega$ in the optical,
and $\sim1.2\times$ stronger in the X-ray), which is the cause of the
numerous beat frequencies visible in the Figure \ref{power}.

Figure \ref{pdm} is a time-sampled thetagram, where the gray scale
indicates $\Theta$ as in Figure \ref{power}.  This plot was made by a
modified version of the PDM code allowing for time-sampled thetagrams
to be produced.  Here we used 5000 data points per segment, and then
stepped through the entire data set by increments of 500 data points
(data point refers to a 10 sec bin and the sets of data points do not
necessarily have equal time sampling).  We thus produced the trailed
thetagram seen in Figure \ref{pdm}, which covers the time period from
2000 May 5 to June 10, the entire {\it EUVE} observation.  Along the
y-axis, one time step refers to a bin of 5000 data points.  The known
periods and sidebands, including the satellite beat frequencies, are
fairly constant and strong throughout the temporal sampling covered in
the plot.  However, there are some interesting frequencies around 25
day$^{-1}$ and 33 day$^{-1}$ which appear suddenly in the middle of
the data set.  These frequencies are likely to be beat frequencies,
but their appearance during only part of the observation is intriguing
and suggests that one or more of the beating frequencies is not
constant throughout the data set, i.e., the geometry and/or physical
conditions of EX Hya are evolving, to some extent.  The frequencies
appearing at $\sim26$ day$^{-1}$ are associated with the frequencies
$\omega/2+\Omega$ and $\omega/2+\Omega_{\rm sat}$ but it is
interesting to note that what appears to be the $\omega/2+\Omega$
frequency wanders back and forth throughout the entire data set.

The appearance of the wandering and intermittent periods in our high
energy data probably indicates that there are additional or changing
sources of reprocessing of the EUV radiation.  These may always be
present in the EUV or may be manifestations of the variable
nature of EX Hya.  A detailed study of the periodicities seen in our
contemporaneous observations of EX Hya in the X-ray, EUV, UV, and
optical will be presented in a future work.

\subsection{EUVE Spectra}\label{sect_spec} 
Figure \ref{spec} presents EUV spectra extracted from the four white
dwarf spin phases, as described in Figure \ref{splc}, and Table
\ref{t2} presents the stronger lines we have identified in each
spectrum, along with their fluxes and $\log\,T_{\rm max}$.  The
emission lines seen in these spectra are from highly ionized neon and
iron, with the coolest lines (neon) having $T_{\rm max} =
10^{5.7}$ K and the hottest lines (iron) having temperatures of
$T_{\rm max} = 10^{7.1}$ K.  Only five lines appear (above the
$2\sigma$ error) in each of the four spin phased spectra, four of
which are blended: \ion{Fe}{15} $\lambda73.47$\AA, \ion{Fe}{20}
$\lambda93.78$\AA/\ion{Fe}{18} $\lambda93.93$\AA, and \ion{Fe}{23}
$\lambda132.84$\AA/\ion{Fe}{20} $\lambda132.85$\AA.  These lines have
$\log\, T_{\rm max}=$ 6.3, 6.9, 6.7, 7.1, and 6.9 K, respectively.
Using the flux ratios of these lines, we can further confine the
temperature range of the emitting plasma.  According to the optically
thin plasma model of Mewe, Gronenschild, \& van den Oord (1985), the
flux ratios of the lines at $\lambda132/\lambda93$\AA,
$\lambda132/\lambda73$\AA, and $\lambda93/\lambda73$\AA\ match
those expected for a $\log T\sim6.5-6.9$ K plasma.  The flux ratio
determined temperature is in agreement with the $T_{\rm max}$ values
of the individual lines.  However, if we consider the flux ratio of
the two Ne lines we have identified in the spin maximum and decline
spectra, \ion{Ne}{8}/\ion{Ne}{6} $\lambda98$\AA\ to \ion{Ne}{8}
$\lambda88$\AA, the plasma temperature determination is significantly
lower, near $\log T\sim5.8$ K, for both spin phases.

Line ratios may also be used to determine plasma densities according
to \citet{mewe85}, Table IVa, using lines that have $\log\,T_{\rm
max}=7.0$ K.  Only two of the four spectra have emission lines near
this temperature which are unblended: spin maximum and spin decline.
The {\it spin maximum} line ratio, \ion{Fe}{20}
$\lambda110.63$\AA/\ion{Fe}{22} $\lambda114.39$\AA$=0.57\pm0.22$,
gives a density of $>10^{14}\cm^{-3}$ (note that \ion{Fe}{20}
$\lambda110.63$\AA\ actually has $\log\,T_{\rm max}=6.9$).  The line
ratio of \ion{Fe}{21} $\lambda117.51$\AA/\ion{Fe}{21}
$\lambda121.21$\AA$=0.80\pm0.72$ in the {\it spin decline} spectrum
gives a lower density, $n_{\rm e}=10^{10}-10^{11}\cm^{-3}$ ($n_{\rm
e}< 10^{13}\cm^{-3}$ when considering the lower limit determined from
the uncertainty).  These values may indicate an observed density
range, although poorly constrained.  We would like to note that the
density determined from the $\lambda110/\lambda114$ line ratio does
have a problem, as the ratio uses lines from two different charge
states and depends strongly on the plasma temperature.  For example,
if we were to initially assume a low density plasma, Table IV of
\citet{mewe85} shows that the value of the $\lambda110/\lambda114$
ratio measured here matches well with a $\log\,T_{\rm max}\sim6.8$ K.
Considering that our spin decline line ratio finds a low density, it
is likely that assuming a constant temperature of $T=10^7$ K for the
$\lambda110/\lambda114$ ratio does not adequately take into account
the strong dependence on temperature of the different charge states,
and that the lines are likely to be formed in a lower density plasma.

Using only the reliably determined density above ($n_{\rm
e}=10^{10}-10^{11}\cm^{-3}$), our value does not agree with the
electron densities which have been previously determined for EX Hya.
\citet{hur97} found an electron density of $n_{\rm e}\geq
10^{13}\cm^{-3}$ using the ratio of the \ion{Fe}{20}/\ion{Fe}{23}
blend to the \ion{Fe}{21} emission lines in their {\it EUVE} spectra.
\citet{mauche01} found a value of $n_{\rm e}\geq 3\times
10^{14}\cm^{-3}$ from our {\it Chandra} spectrum using the
\ion{Fe}{17} $\lambda17.10/\lambda17.05$\AA\ line ratio.  We note that
while these two values are self-consistent, the \citet{hur97} value
was determined from blended lines in different charge states and the
\citet{mauche01} ratio is affected by photoexcitation.

The appearance of the \ion{Fe}{23}/\ion{Fe}{20} lines at
$\lambda132$\AA\ in each spectrum (except during mid-bulge dip, see
Figure 2 of Belle et al. 2002b) suggests that a high temperature
plasma ($10^7$ K) is constantly in view during the times in which we
see other EUV emitting regions of lower temperature and lower density.
The absence of the $\lambda132$\AA\ line, and also the $\lambda93$\AA\
line, during the mid-bulge dip phase suggests that these lines are
formed interior to the bulge and that their emitting region is roughly
confined to the orbital plane.

The emission line centers appear to shift over the four spin phases.
Each spin phase is comprised of many orbital phases, so radial
velocities of these lines will be a complex sum of both motions.
Also, a radial velocity curve over spin phase will not have much
meaning when only containing four points.  However, it is interesting
to note that the \ion{Fe}{23}/\ion{Fe}{20} blend at $\lambda132$\AA\
has a fairly constant negative velocity at all spin phases with an
average value of $-260\pm180\kms$.  The other two strong lines which
appear in each of the four spectra also have negative velocities over
all four spin phases.  The \ion{Fe}{15} $\lambda73.47$\AA\ emission
line has its most blue-shifted velocity of $-980\pm245\kms$ during
spin maximum and its lowest blue-shifted velocity of $-531\pm572\kms$
during spin decline.  The \ion{Fe}{20}/\ion{Fe}{18} blend at
$\lambda93$\AA\ is not blue-shifted to the same extent and has its
greatest blue-shift of $-463\pm256\kms$ during spin minimum.  The SW
spectrometer rms error of $0.11$\AA\ \citep{abbott96}, which has not
been added to the $1\sigma$ errors given for the line velocities
above, translates to a velocity error of $450\kms$ at $73$\AA,
$350\kms$ at $93$\AA, and $250\kms$ at $132$\AA; even assigning a
systematic maximum uncertainty to the line velocities, we note that
the emission lines are generally still blue-shifted.  (The
$\gamma-$velocity of EX Hya has not been subtracted from these values
due to the fact that it is not well constrained.)

Taken at face value, the phase-independent blue-shifts of the emission
lines in the {\it EUVE} spectra tell us that the lines may be
associated with some outflowing material.  However, we note that
\citet{mauche01} reported that the \ion{Fe}{17} $\lambda15-17$\AA\
emission lines had a wavelength offset consistent with zero (though
the error from the wavelength accuracy they quote does allow these
lines to be blue-shifted.)  The accretion pole regions would be a
likely site for outflowing material in the form of a wind or jet.  The
disappearance of the $\lambda132$\AA\ line during the bulge dip would
mean that the poles must be highly inclined with respect to the spin
axis if the cause of its blue-shift is due to jets.  We believe
instead that this outflow may originate as a wind from the inner edge
of the accretion disk, a feature common in high mass transfer
binaries, possibly associated with shocks in the accretion disk (see
below).  Such an emission site would constantly provide outflowing
material directed at the observer, while also being absorbed during
the bulge dip phases.  In an upcoming paper discussing our {\it HST}
data \citep{bel02a}, we show that blue-shifted absorption components
(P Cygni profiles), features which are indicative of outflowing
material, are seen in several of the UV emission lines.

\section{A REVISED MODEL FOR EX HYA}
The current intermediate polar model has the white dwarf evenly
accreting matter from an annular accretion disk via accretion curtains
onto both magnetic poles (e.g., Patterson 1994).  However,
\citet{king99} have recently shown that EX Hya exists in a spin
equilibrium state such that this model may not hold true.  They first
note that although a disk may form in any intermediate polar, it is
generally far from Keplerian due to the influence of the magnetic
field on the accretion flow in the disk.  For EX Hya, where the spin
period is close to the binary orbital period ($P_{\rm spin}/P_{\rm
orb}=0.68$ for EX Hya compared with $P_{\rm spin}/P_{\rm orb}\sim0.1$
for other IPs), the equilibrium condition is such that the corotation
and the circularization radii are roughly equal to the distance to the
inner Lagrange point $L_1$.  This condition is true for a special
class of intermediate polars which possess a white dwarf with a
magnetic moment of $2\times10^{33}\mo\simlt\mu_1\simlt10^{34}\mo$ (and
$P_{\rm orb}<2$ hr), corresponding to a magnetic field strength of
$2-7$ MG for the white dwarf in EX Hya.

The EUV data presented in this paper support the model alluded to by
the \citet{king99} study in which non-uniform accretion from the disk
onto the white dwarf magnetic poles occurs.  The extended bulge dip
seen in the binary light curve (Figure \ref{bplc}), the absorption
dips seen throughout the binary light curves of Figure
\ref{wdcompare}, along with the differing absorption amounts seen
between these same light curves are all evidence of non-uniform
accretion.  We propose that the magnetic field in EX Hya is strong
enough to begin pulling some material vertically from the outer edge
of the disk, thereby forming a large accretion curtain which would
produce the very broad and extended bulge.  Material after binary
phase 0 would accumulate as it encounters the bulge in the accretion
disk.  This concept is depicted in Figure \ref{curtain} (although we
have drawn accretion curtains which extend only from the inner or
outer disk, it is more likely that there would be a continuous curtain
extending between the two disk edges.)  Matter attaching to the
magnetic field lines at the outer edge of the disk would form
accretion streams which would act to absorb radiation emitted from the
accretion poles and would produce dips within the binary light curves,
such as those seen at $\pb=0.97$ and $\pb=1.04$.  These dips are
reminiscent of dips seen in polar light curves (e.g., Belle, Howell,
\& Mills 2000).  Inspection of Figure \ref{wdcompare} shows that the
dips at $\pb=0.97$ and $\pb=1.04$ have varying depths (values are
given in Table \ref{countrate}), with the dips during spin minimum
having the greatest amount of absorption.  At spin and binary phases
such that spin minimum and the bulge eclipse are coincident
($\ps=0.37-0.87$ and $\pb=0.6-1.0$), material in the bulge, already
extending above the plane of the disk, may be controlled by the
magnetic field and produce denser streams.

If the dips seen in the binary phased light curve are caused by
material attached to the field lines, why are the dips seen along the
binary phase rather than the spin phase?  The reality of the magnetic
field attachment to the outer edge of the disk is that the field lines
will be affected by magnetic torque such that the lines will become
twisted.  While the white dwarf magnetic pole has already started to
rotate away from the bulge, the distortion of the magnetic field
caused by the torque will allow the field lines to remain attached at
the bulge.  As long as the field lines remain attached, matter will be
kept from falling back down to the accretion disk.  In this way, the
dips would be seen along the binary phase rather than the spin phase.

If the scenario just described is true for EX Hya, the strength of the
magnetic field needed to pull material from the outer edge of the disk
can be estimated.  Using the equation,
\begin{equation}
r_{\mu}=9.9\times10^{10}\,\mu_{34}^{4/7}\,M_{\rm
WD}^{-1/7}\,\dot{M}_{16}^{-2/7}\cm
\end{equation}
to calculate the radius of the magnetosphere \citep{warner95} and
setting $r_{\mu}$ equal to the outer accretion disk radius, we can
solve for $\mu_{34}$.  We use white dwarf masses of $0.48\msun$
\citep{fuj97} and $0.85\msun$ \citep{hur97} to represent the range of
primary masses which have been determined for EX Hya.  Optical
spectroscopic emission line velocities (our optical data will appear
in a future paper), and the white dwarf mass range, give the outer
disk radius as $1.6\times10^{10}$ or $4.3\times10^{10}\cm$.  With an
accretion rate of $1\times10^{16}\,\gs$ \citep{hur97}, we find a
magnetic field strength of $\sim1-8$ MG.  We would like to note that
using the optical emission line velocities to determine the disk
radius assumes that some of the disk is in Keplerian motion.

The far-reaching magnetic field has implications other than producing
an extended bulge.  As stated previously, the circularization radius
is roughly equal to the corotation radius \citep{king99}.  This
intimates that part or possibly all of the accretion disk is rotating
with the white dwarf.  Although we do not know the latitude of the
magnetic poles on the white dwarf surface, the modulation seen on the
spin period suggests that the poles are moderately inclined
($\sim10^{\circ}$, Kim \& Beuermann 1995, 1996) with respect to the
rotation axis of the white dwarf.  The accretion disk, then, would not
feel a constant magnetic force, but rather would feel the highest
field strength at the regions of the disk closest to the poles.  At
these locations along the disk, the disk material would be more
affected by the magnetic field strength and hence would become
controlled by the magnetic field.  Thus, we have a situation in which
some amount of material in two chunks, or pie slices (one facing each
pole), along the accretion disk are rotating with the white dwarf,
while the rest of the accretion disk may (or may not) be rotating with
the white dwarf.  The outer disk velocities measured from the optical
spectra are around $500-600\kms$, while the velocity of the magnetic
field sweeping through the disk is calculated at the corotation radius
($R_{\rm co}=3.5\times10^{10}$ cm) to be $\sim550\kms$, again
enforcing the statement that the outer disk is corotating with the
white dwarf magnetic field.

An issue pertaining to the corotating pie slices in the accretion disk
is the presence of shocks which may form as matter of different
velocities interacts in the accretion disk.  Modeling shocks is beyond
the scope of this paper, however, we argue that the similar velocities
of the white dwarf and accretion disk (as discussed previously) would
produce very low energy shocks.  High energy emission lines are
already seen in the spectra and because the bulge dip eclipses all EUV
emission (i.e., we cannot distinguish between the white dwarf or
accretion disk being eclipsed), it is not possible to discern if these
lines originate in a shock from the accretion disk, or from the shocks
and high energy plasmas associated with regions near the magnetic
accretion poles.  However, if the shocks produce winds, they may be
responsible for the highly blue-shifted lines seen in the EUV spectra.

Another question to answer is why strong power is not seen at the
$\omega-\Omega$ frequency if the magnetic field is coupled to the
accretion disk (though we note that while this frequency is not
strong, it is above the 95\% confidence level.)  This might be
expected if there was a bright EUV point source situated on the edge
of the disk, however, the outer edge of the disk is vertically
extended and cool and will emit most strongly at lower energies, such
as optical wavelengths.  We will show in a future paper describing our
optical data that strong power {\em is seen} on the $\omega-\Omega$
frequency.  We should also consider the possibility that an overflow
accretion stream may be responsible for the small dips seen at
$\pb=0.97$ and $\pb=1.04$, instead of magnetically controlled streams
causing these dips.  We argue that this is not likely for several
reasons.  First, there are many dips seen throughout the bulge dip,
starting at $\pb\sim0.6$ and continuing through to $\pb\sim0$, which
makes it likely that these dips are caused by material associated with
the bulge.  Second, the overflow stream is unlikely to extend far
enough above the orbital plane in order to be along the line of sight
to the accretion poles.  Third, this overflow stream would follow
along its ballistic trajectory and would not be in a position to cause
a dip occurring after binary phase 0.

\section{CONCLUSION}
\begin{figure}
\begin{center}
%\epsscale{0.8} 
\plotone{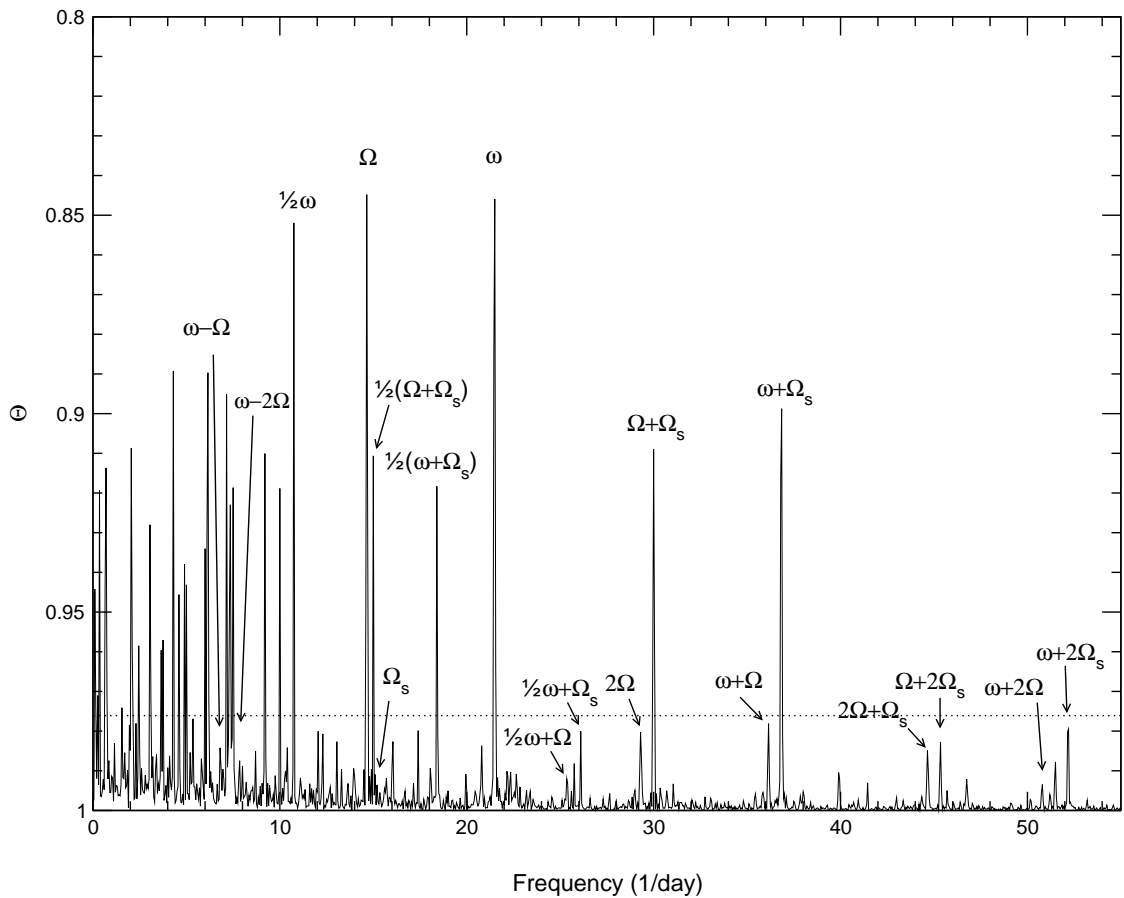} 
\figcaption[]{Thetagram of the EUV photometric data set.  The known
optical orbital sidebands are noticeable and have been labeled, where
$\omega$ represents the spin frequency, $\Omega$ represents the binary
frequency, and $\Omega_{\rm s}$ is the satellite orbital frequency.
We have also labeled several other strong frequencies which we discuss
in the text.  The dashed line represents the 99\% confidence level at
0.976 in power and the 95\% confidence level is near 0.99 in power.
The frequency resolution is 0.05.
\label{power}}
\end{center}
\end{figure}

\begin{figure}
\begin{center}
\epsscale{0.7}
\plotone{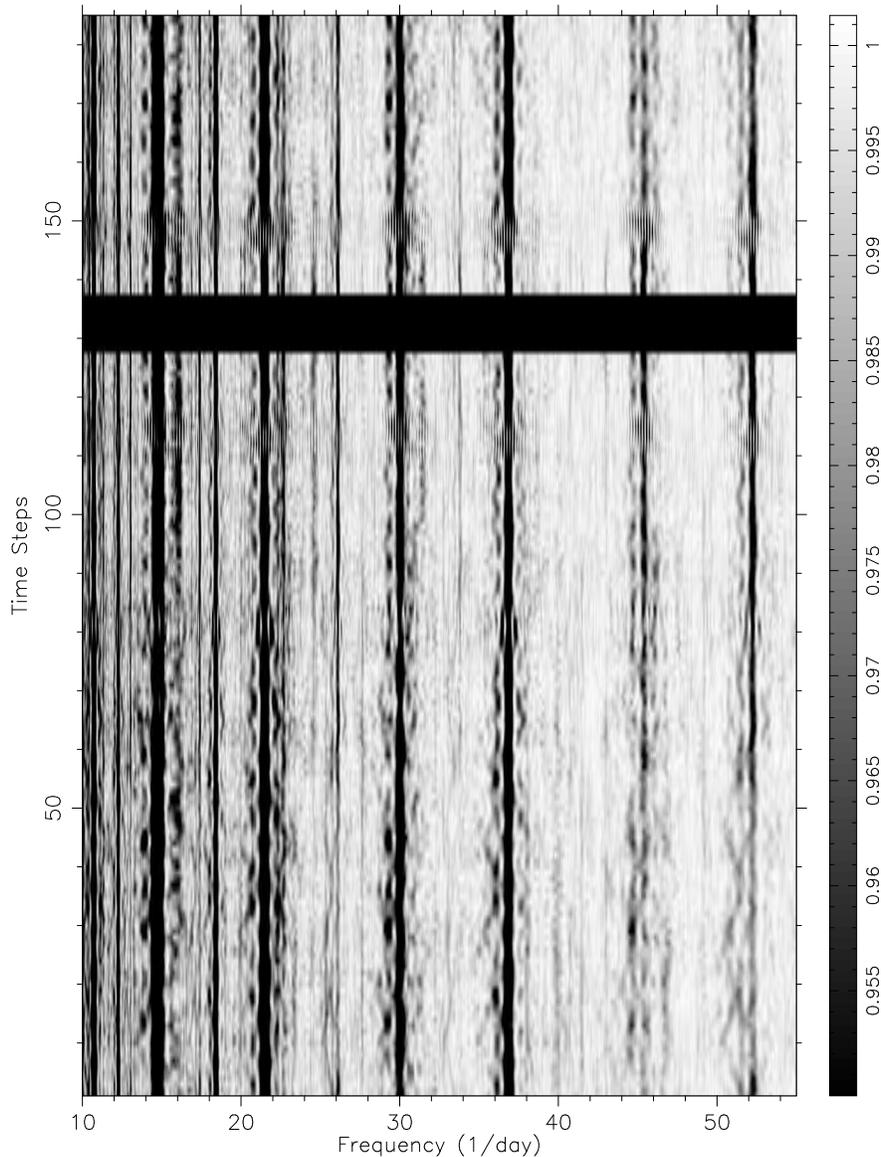}
\figcaption[]{A time series thetagram of the EUV photometric data.
The x-axis covers the frequency range $10-55\;{\rm day}^{-1}$ while
the y-axis is time from the beginning to end of the {\it EUVE}
observation.  The gray scale indicates power, with darker areas
corresponding to higher confidence levels; the color bar has the same
numeric meaning as the y-scale in Figure \ref{power}.  The more
powerful frequencies seen in Figure \ref{power} are also apparent
here, and appear to remain constant throughout the time series, while
less powerful frequencies come and go (e.g., see near 25 and 33
day$^{-1}$).  The region of missing data is due to a number of days
with poor temporal sampling caused by an increased background level in
the detector.
\label{pdm}}
\end{center}
\end{figure}

\begin{figure}
\begin{center}
\plotone{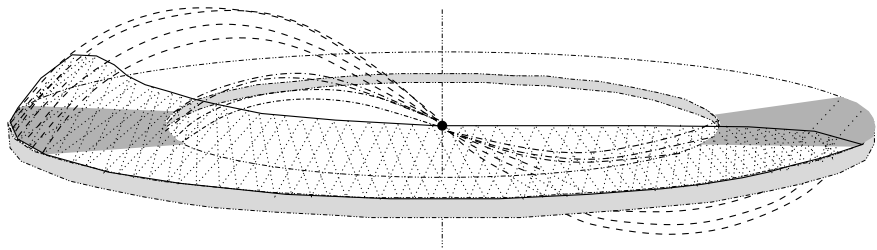}
\figcaption[]{A schematic view of the inner accretion region in EX
Hya.  The parts of the accretion disk which may be corotating with the
field lines are indicated (dark gray regions) and the extended bulge
has been drawn (hatched region extending along the outer edge of the
disk).  The size of these regions are estimated.  The dashed lines
extending from the white dwarf to the inner and outer edges of the
accretion disk represent the inner and outer accretion curtains,
respectively. The view of EX Hya drawn here corresponds to binary
phase $\pb=0.75$ and spin phase $\ps=0.75$.  The bulge would also
extend below the orbital plane.
\label{curtain}}
\end{center}
\end{figure}

The photometric and spectroscopic data presented here have provided us
with some intriguing results for EX Hya.  This system clearly does not
conform to the traditional intermediate polar model.  A broad bulge
dip, numerous smaller dips throughout the binary light curve, and the
inconsistencies between binary light curves extracted at different
spin phases all indicate that EX Hya may be accreting material from
the outer edge of the accretion disk thereby producing an extended
bulge and large accretion curtain.  For material at the outer edge of
the disk to become magnetically controlled, the magnetic field of the
white dwarf must be on the order of $1-8$ MG.  These extended regions
may also be reprocessing sites of EUV radiation, as can be seen from
the many strong beat frequencies appearing in the single thetagram
(Figure \ref{power}) and time-series thetagram (Figure \ref{pdm}).
Iron spectral line ratios give a temperature of $10^{6.5}-10^{6.9}$ K
for the emitting plasma but provide only a poorly constrained density
of $n_{\rm e}=10^{10}-10^{11}\cm^{-3}$.  It is likely that regions of
varying temperature and density are viewed simultaneously throughout
the spin orbit.  Future papers will investigate the physical meaning
of the strong frequencies in greater detail, along with relating the
EUV data to our other wavelength observations.

\acknowledgments 
We would like to thank C. Mauche for his contributions and comments on
this work and the anonymous referee for very valuable comments.  This
research was partially supported by NASA grant NAG5-8644 to SBH.

\newpage

\begin{deluxetable}{lcccccccc}
\tabletypesize{\footnotesize} 
\rotate
\tablewidth{0pt}
\tablecolumns{9}
\tablecaption{Identified bright EUV spectral emission lines. \label{t2}}
\tablehead{
\colhead{} & \multicolumn{8}{c}{Spin Phase} \\
\colhead{} & \multicolumn{2}{c}{Maximum} & \multicolumn{2}{c}{Decline} & 
\multicolumn{2}{c}{Minimum} & \multicolumn{2}{c}{Rise} \\ 
\colhead{Ion $\lambda$ ($\log\,T_{\rm max}$)} & 
\colhead{$\lambda_{\rm obs}$\tablenotemark{a}} & 
\colhead{Flux\tablenotemark{b}} &
\colhead{$\lambda_{\rm obs}$} & \colhead{Flux} &
\colhead{$\lambda_{\rm obs}$} & \colhead{Flux} &
\colhead{$\lambda_{\rm obs}$} & \colhead{Flux} \\
\colhead{(\AA) (K)} & 
\colhead{(\AA)} & \colhead{} &
\colhead{(\AA)} & \colhead{} &
\colhead{(\AA)} & \colhead{} &
\colhead{(\AA)} & \colhead{}}  
\startdata
\ion{Fe}{15} $73.47$ (6.3) & $73.23\pm0.06$ & $11.9\pm3.60$ 
			& $73.34\pm0.14$ & $8.04\pm5.18$
                         & $73.30\pm0.06$ & $10.3\pm3.44$ 
			& $73.27\pm0.05$ &$12.1\pm4.14$\\

%\ion{Ne}{9}  $74.53$ (6.6) & $74.34$ & $4.04\pm0.06$ & \nodata & \nodata  
%                         & \nodata & \nodata & \nodata & \nodata   \\

%\ion{Ne}{8} $74.60$ (5.8) \\ \ion{Fe}{13} $74.85$ (6.2) &
%                              $74.62$ & $2.20\pm0.01$ & \nodata & \nodata
%		         & $74.75$  &$4.20\pm0.20$ & $74.69$ & $4.77\pm0.12$\\

%\ion{Ne}{7} $75.77$ (5.8) \\ \ion{Fe}{13} $75.89$ (6.2) &
%                              $75.52$ & $3.90\pm0.10$ & \nodata & \nodata 
%                         & $75.78$ &$4.99\pm0.38$ & $75.74$ & $4.92\pm0.07$ \\

%\ion{Fe}{12} $80.02$ (6.2)  & $80.07$ & $1.77\pm0.02$ & \nodata & \nodata  
%		          & $80.08$ & $2.17\pm0.05$ & \nodata & \nodata \\

%\ion{Fe}{12} $80.51$ (6.2)  & $80.43$ & $1.82\pm0.14$ & \nodata & \nodata 
%                          & \nodata & \nodata & \nodata & \nodata   \\

%\ion{Ne}{7}  $84.29$ (5.8)  & $84.06$ & $2.60\pm0.20$ & \nodata & \nodata 
%                          & \nodata & \nodata  & \nodata& \nodata   \\

\ion{Ne}{8}  $88.08$ (5.8)  & $88.03\pm0.26$ & $3.75\pm3.29$
  			& $87.95\pm0.08$ & $1.97\pm0.91$
                          & $88.00\pm0.08$ & $1.43\pm0.60$ 
			& \nodata & \nodata   \\

%\ion{Fe}{14} $91.01$ (6.3) \\ \ion{Fe}{19} $91.02$ (6.8) &
%                           $90.77$ & $5.00\pm0.80$ & $90.84$ & $1.79\pm0.10$
%			&  \nodata & \nodata & \nodata & \nodata \\

\ion{Fe}{20} $93.78$ (6.9)/ \\ $~~~$\ion{Fe}{18} $93.93$ (6.7)   &
                            $93.83\pm0.04$ & $5.28\pm0.73$ 
			& $93.85\pm0.06$ & $4.94\pm0.85$
			 & $93.71\pm0.08$ & $2.16\pm0.62$ 
			& $93.78\pm0.07$ & $3.53\pm0.98$\\

%\ion{Ne}{7} $97.50$ (5.7) \\ \ion{Fe}{21} $97.87$ (7.0)  & 
%		              \nodata & \nodata & $97.81$ & $2.36\pm0.35$  
%                         & $97.55$ & $7.79\pm0.12$  & \nodata & \nodata \\

\ion{Ne}{8} $98.13$ (5.8)/ \\ $~~~$\ion{Ne}{6} $98.20$ (5.7)   &
                         $98.16\pm0.06$ & $6.13\pm1.08$
			& $98.33\pm0.12$& $3.23\pm1.27$ 
 			& \nodata & \nodata & \nodata & \nodata  \\

%\ion{Fe}{19} $101.55$ (6.8) & \nodata & \nodata & $101.50$ & $1.84\pm0.16$ 
%			& \nodata & \nodata & \nodata  & \nodata  \\

\ion{Fe}{21} $102.21$ (7.0)/ \\ $~~~$\ion{Fe}{22} $102.23$ (7.0)  &
                             $102.21\pm0.05$ & $2.48\pm0.76$ 
			& \nodata& \nodata 
			& \nodata & \nodata  
			& $102.20\pm0.10$ & $2.71\pm1.05$ \\

%\ion{Ne}{7} $106.09/106.19$ (5.7) &
%                         $105.90$ & $2.40\pm0.15$ & $105.90$ & $1.74\pm0.07$ 
%			& \nodata & \nodata & \nodata  & \nodata     \\

%\ion{Fe}{21} $108.37$ (7.0)/ \\ $~~~$\ion{Fe}{19} $108.37$ (6.8) &
%                         $108.17$ & $2.71\pm0.53$ & $108.21$& $1.64\pm0.05$ 
%			& \nodata & \nodata & $108.16$ & $2.67\pm0.17$  \\

\ion{Fe}{20} $110.63$ (6.9) & $110.62\pm0.06$ & $2.16\pm0.67$ 
			& \nodata & \nodata 
			& \nodata & \nodata 
			& $110.77\pm0.10$ & $2.41\pm0.96$ \\

\ion{Fe}{21} $113.30$ (7.0)/ \\ $~~~$\ion{Fe}{20} $113.34$ (6.9) &
                        $113.26\pm0.06$ & $3.90\pm0.71$ 
			& $113.20\pm0.09$ & $2.17\pm0.78$ 
			& \nodata & \nodata & \nodata  & \nodata  \\

\ion{Fe}{22} $114.39$ (7.0) & $114.28\pm0.05$ & $3.80\pm0.85$ 
			& \nodata & \nodata  
			& \nodata & \nodata & \nodata  & \nodata  \\

\ion{Fe}{22} $117.17$ (7.0) & $117.07\pm0.05$ & $3.41\pm0.78$ 
			& $117.09\pm0.07$ & $1.87\pm0.80$ 
			& \nodata & \nodata & \nodata  & \nodata  \\

\ion{Fe}{21} $117.51$ (7.0) & \nodata & \nodata 
			& $117.70\pm0.04$ & $0.99\pm0.51$ 
			& \nodata  &\nodata & \nodata  & \nodata   \\

%\ion{Fe}{19} $120.00$ (6.8) & \nodata & \nodata & $120.14$ & $1.10\pm0.03$ 
%			& \nodata & \nodata  & $120.20$ & $2.29\pm0.24$ \\

\ion{Fe}{21} $121.21$ (7.0) & \nodata & \nodata 
			& $121.19\pm0.10$ & $1.24\pm0.92$ 
			& \nodata & \nodata & \nodata  & \nodata  \\

\ion{Fe}{23} $132.84$ (7.1)/ \\ $~~~$\ion{Fe}{20} $132.85$ (6.9)  &
                        $132.74\pm0.03$ & $9.11\pm1.48$ 
			& $132.72\pm0.04$ & $5.01\pm1.23$
	           & $132.71\pm0.06$ & $2.02\pm0.95$ 
			& $132.74\pm0.07$ & $4.55\pm1.67$\\
\enddata
\tablenotetext{a}{Errors on all $\lambda_{\rm obs}$ measurements are
$1\sigma$ errors and do not include the rms error of the SW spectrometer.}
\tablenotetext{b}{Fluxes are in units of $10^{-4}$ photons s$^{-1}$ 
cm$^{-2}$.}
\end{deluxetable}

\begin{deluxetable}{ccc}
\tablecolumns{3}
\tablewidth{0pc}
\tablecaption{Binary light curve mid-dip count rates.  \label{countrate}}
\tablehead{
\colhead{Spin Phase} & \colhead{$\pb=0.97$ (\% max)\tablenotemark{a}} & 
\colhead{$\pb=1.04$ (\% max)}\\
\colhead{} & \colhead{(counts s$^{-1}$)} & \colhead{(counts s$^{-1}$)} }
\startdata
Maximum & $0.070\pm0.009$ ($-$)   &  $0.089\pm0.010$ ($-$)\\ 
Minimum & $0.003\pm0.004$ (5)     &  $0.011\pm0.005$ (13) \\
Rise    & $0.018\pm0.007$ (25)    &  $0.035\pm0.009$ (40) \\
Decline & $0.032\pm0.007$ (45)    &  $0.055\pm0.008$ (62) \\
\enddata
\tablenotetext{a}{``\% max'' indicates the dip count rate as a
percentage of the maximum mid-dip count rate.}
\end{deluxetable}

\begin{deluxetable}{ccccccc}
\tablecolumns{9}
\tablewidth{0pc}
\tablecaption{Gaussian fit results for the two strongest photometric dips.
\label{times}}
\tablehead{
\colhead{} & \multicolumn{3}{c}{$\pb=0.97$} 
& \multicolumn{3}{c}{$\pb=1.04$} \\
\colhead{Spin Phase} & \colhead{Midpoint\tablenotemark{a}} & \colhead{FWHM}
& \colhead{FWZI} & \colhead{Midpoint} & \colhead{FWHM} & \colhead{FWZI}\\
\colhead{} & \colhead{($\pb$)} & \colhead{($\Delta\pb$)} & 
\colhead{($\Delta\pb$)}  
& \colhead{($\pb$)} & \colhead{($\Delta\pb$)} & \colhead{($\Delta\pb$)}  }
\startdata
Averaged& $0.970$ & $0.010\pm0.001$ & 0.024
	& $1.037$ & $0.018\pm0.002$ & 0.044 \\

Maximum & $0.969$ & $0.012\pm0.001$ & 0.019
	& $1.038$ & $0.018\pm0.002$ & 0.041 \\

Minimum & $0.972$ & $0.014\pm0.001$ & 0.025
	& $1.037$ & $0.018\pm0.002$ & 0.031 \\

Rise    & $0.969$ & $0.010\pm0.001$ & 0.030
	& $1.038$ & $0.022\pm0.001$ & 0.045 \\

Decline & $0.972$ & $0.016\pm0.001$ & 0.030 & $1.037$ &
	$0.012\pm0.001$ & 0.025 \\ 
\enddata 
\tablenotetext{a}{Errors on the midpoint values fall within the
resolution of the light curves.}
\end{deluxetable}

%\begin{deluxetable}{ccc}
%\tablecolumns{3}
%\tablewidth{0pc}
%\tablecaption{New orbital sideband frequencies.
%\label{freq}}
%\tablehead{ 
%\colhead{} & \colhead{Frequency} & \colhead{Period} \\
%\colhead{Number} & \colhead{(day$^{-1}$)} & \colhead{(min)}} 
%\startdata 
%1 & $4.294\pm0.023$ & 335.352 \\ 
%2 & $10.744\pm0.045$ & 134.028 \\ 
%3 & $18.418\pm0.010$ & 78.184 \\ 
%4 \tablenotemark{a} & $30.002\pm0.089$ & 47.996 \\ 
%5 \tablenotemark{a} & $36.831\pm0.092$ & 39.097 \\ 
%\enddata 
%\tablenotetext{a}{We note that while these two frequencies are close
%to the orbital sidebands of $\Omega+\Omega_{\rm sat}$ (4) and $\omega+
%\Omega_{\rm sat}$ (5), inspection of a high resolution power spectrum
%shows that they are not the same.}
%\end{deluxetable}

\newpage

\begin{thebibliography}{99}
%
\bibitem[Abbott et al.(1996)]{abbott96}
Abbott, M. J. et al.
1996, \apjs, 107, 451
%
\bibitem[Allan et al.(1998)]{allan98}
Allan, A., Hellier, C., \& Beardmore, A.
1998, \mnras, 295, 167
%
\bibitem[Belle et al.(2002a)]{bel02a}
Belle, K. E., et al.
2002a, in preparation
%
\bibitem[Belle et al.(2000)]{belle00}
Belle, K. E., Howell, S. B., \& Mills, A.
2000, \pasp, 112, 343
%
\bibitem[Belle et al.(2002b)]{belle02}
Belle, K. E., Howell, S. B., \& Sirk, M.
2002b, in ASP Conf. Ser. 264, Continuing the Challenge of 
EUV Astronomy: Current Analysis and Prospects for the Future, 
ed. S. Howell, J. Dupuis, D. Golombek, F. Walter, \& J. Cullison 
(San Francisco: ASP), 96
%
\bibitem[Beuermann \& Osborne(1988)]{beuer88}
Beuermann, K. \& Osborne, J. P.
1988, \aap, 189, 128
%
\bibitem[Bowyer(1995)]{bowyer95}
Bowyer, S.
1995, Proc. SPIE, 2517, 97
%
\bibitem[Cordova et al.(1985)]{cord85}
C\'{o}rdova, F. A., Mason, K. O., \& Kahn, S. M.
1985, \mnras, 212, 447
%
\bibitem[Fujimoto \& Ishida(1997)]{fuj97}
Fujimoto, R. \& Ishida, M.
1997, \apj, 474, 774
%
\bibitem[Hellier(1997)]{hellier97}
Hellier, C.
1997, \mnras, 291, 71
%
\bibitem[Hellier et al.(2000)]{hellier00}
Hellier, C., Kemp, J., Naylor, T., Bateson, F. M., Jones, A., 
Overbeek, D., Stubbings, R., \& Mukai, K.
2000, \mnras, 313, 703
%
\bibitem[Hellier et al.(1987)]{hellier87}
Hellier, C., Mason, K. O., Rosen, S. R., \& C\'{o}rdova, F. A.
1987, \mnras, 228, 463
%
\bibitem[Hellier \& Sproats(1992)]{ephem}
Hellier, C. \& Sproats, L. N.
1992, IBVS, 3724
%
\bibitem[Hurwitz et al.(1997)]{hur97}
Hurwitz, M., Sirk, M., Bowyer, S., \& Ko, Y.-K.
1997, \apj, 477, 390
%
\bibitem[Kaitchuck et al.(1987)]{kait87}
Kaitchuck, R. H., Hantzios, P. A., Kakaletris, P., Honeycutt, R. K., \&
Schlegel, E. M.
1987, \apj, 317, 765
%
\bibitem[King \& Wynn(1999)]{king99}
King, A. R. \& Wynn, G. A.
1999, \mnras, 310, 203
%
\bibitem[Kim \& Beuermann(1995)]{kim95}
Kim, Y. \& Beuermann, K.
1995, \aap, 298, 165
%
\bibitem[Kim \& Beuermann(1996)]{kim96}
Kim, Y. \& Beuermann, K.
1996, \aap, 307, 824
%
\bibitem[Mason et al.(2000)]{mason00}
Mason, E. et al.
2000, \mnras, 318, 440
%
\bibitem[Mauche(1999)]{mauche99}
Mauche, C. W.
1999, \apj, 520, 822
%
\bibitem[Mauche, Liedhal, \& Fournier(2001)]{mauche01}
Mauche, C. W., Liedhal, D. A., \& Fournier, K. B.
2001, \apj, 560, 992
%
\bibitem[Mewe et al.(1985)]{mewe85}
Mewe, R., Gronenschild, E. H. B. M., \& van den Oord, G. H. J.
1985, \aaps, 62, 197
%
\bibitem[Mukai et al.(1998)]{mukai98}
Mukai, K., Ishida, M., Osborne, J., Rosen, S., \& Stavroyiannopoulos, D.
1998, in ASP Conf. Ser. 137, Wild Stars in the Old West, 
ed. S. Howell, E. Kuulkers, C. Woodward 
(San Francisco: ASP), 554
%
\bibitem[Mumford(1967)]{mum67}
Mumford, G. S.
1967, \apjs, 15, 1
%
\bibitem[Patterson(1994)]{pat94}
Patterson, J.
1994, \pasp, 106, 209
%
\bibitem[Reinsch \& Beuermann(1990)]{reinsch90}
Reinsch, K. \& Beuermann, K.
1990, \aap, 240, 360
%
\bibitem[Rosen et al.(1988)]{rosen88}
Rosen, S. R., Mason, K. O., \& C\'{o}rdova, F. A.
1988, \mnras, 231, 549
%
\bibitem[Rosen et al.(1991)]{rosen91}
Rosen, S. R., Mason, K. O., Mukai, K., \& Williams, O. R.
1991, \mnras, 249, 417
%
\bibitem[Siegel et al.(1989)]{siegel89}
Siegel, N., Reinsch, K., Beuermann, K., van der Woerd, H., \&
Wolff, E.
1989, \aap, 225, 97
%
\bibitem[Singh \& Swank(1993)]{singh93}
Singh, J. \& Swank, J.
1993, \mnras, 262, 1000
%
\bibitem[Sirk et al.(1997)]{sirk97}
Sirk, M. M., Vallerga, J. V., Finley, D. S., Jelinsky, P., \& Malina R. F.
1997, \apjs, 110, 347 
%
\bibitem[Stellingwerf(1978)]{stell78}
Stellingwerf, R. F.
1978, \apj, 224, 953
%
\bibitem[Warner(1995)]{warner95}
Warner, B.
1995, Cataclysmic Variable Stars (Cambridge: Cambridge Univ. Press)
%
\bibitem[Wood et al.(1986)]{wood86}
Wood, J., Horne, K., Berriman, G., Wade, R., O'Donoghue, D., \& Warner, B.
1986, \mnras, 219, 629
%
\end{thebibliography}
\end{document}